# Intrinsic Rippling Enhances Static Non-Reciprocity in Graphene Metamaterials

Duc Tam Ho[1], Harold Park[2] and Sung Youb Kim[1, *]

[1]Department of Mechanical Engineering, Ulsan National Institute of Science and Technology, Ulsan 44919, South Korea.

[2]Department of Mechanical Engineering, Boston University, Boston, MA 02215, USA.

(2017.07.31)

[*]To whom correspondence should be addressed, e-mail: sykim@unist.ac.kr

**Abstract:** In mechanical systems, Maxwell-Betti reciprocity means that the displacement at point B in response to a force at point A is the same as the displacement at point A in response to the same force applied at point B. Because the notion of reciprocity is general, fundamental, and is operant for other physical systems like electromagnetics, acoustics, and optics, there is significant interest in understanding systems that are not reciprocal, or exhibit non-reciprocity. However, most studies of non-reciprocity have occurred in bulk-scale structures for dynamic problems involving time reversal symmetry.  As a result, little is known about the mechanisms governing static non-reciprocal responses, particularly in atomically-thin two-dimensional materials like graphene.  Here, we use classical atomistic simulations to demonstrate that out of plane ripples, which are intrinsic to graphene, enable significant, multiple orders of magnitude enhancements in the statically non-reciprocal response of graphene metamaterials. Specifically, we find that a striking interplay between the ripples and the stress fields that are induced in the metamaterials due to their geometry impact the displacements that are transmitted by the metamaterial, thus leading to a significantly enhanced static non-reciprocal response. This study thus demonstrates the potential of two-dimensional mechanical metamaterials for symmetry-breaking applications.

**Introduction:** Reciprocity is a fundamental physical principle that has significant implications for a range of scientific disciplines. Essentially, reciprocity implies that the response of a structure at point B to an excitation at point A will be the same as the response of the structure at point A to an excitation at point B, and that this is independent of variations in geometry or material properties. In structural mechanics, this is formalized through the Maxwell-Betti reciprocity theorem, which states that the displacement of point B due to a force at point A is the same as the displacement of point A due to a force at point B. Mathematically, this is written as

$$F_A u_{B \to A} = F_B u_{A \to B}$$

where $F_A$ is the force applied to point A, and $u_{A \to B}$ is displacement of point B induced by $F_A$. Because reciprocity is a general, and fundamental physical principle with applications across the scientific spectrum in electromagnetism, optics, acoustics and mechanics, there has been significant recent interest in systems that break reciprocity, or are non-reciprocal[1–4]. In mechanical systems, this has predominately been for dynamic problems involving wave propagation to break time-reversal symmetry[1–9]. Breaking time-reversal symmetry would enable novel applications and functionality by controlling the direction of wave propagation, enabling filtering and isolation, preventing information backscattering like echoing, and acoustic amplification.

In contrast, while many mechanical systems operate within the static regime, there have been significantly fewer studies on static non-reciprocity in mechanical systems. Recently, Coulais and co-workers performed a seminal study investigating static non-reciprocity in mechanical structures[10]. By combining large nonlinearities and geometric asymmetry, they were able to induce non-reciprocal deformations in both a fishbone structure as well as a mechanical metamaterial. The observation of non-reciprocity in a metamaterial is important both due to the increasing interest in mechanical metamaterials, as well as the recent reports of novel functionality and properties they have been found to exhibit[11–16].

However, nearly all reported studies of non-reciprocal behavior have been for bulk-scale structures, and so with the exception of one report we are aware of on signal isolation in multilayer graphene nanoribbons[5], there is little understanding of the mechanisms governing static non-reciprocity at the nanoscale, and specifically for two-dimensional (2D) mechanical metamaterials. In this work, we investigate the potential of graphene, the canonical 2D material, as a statically non-reciprocal mechanical metamaterial. Graphene is in many ways an ideal material for studies of non-reciprocity, due to its ability to undergo large, geometric shape changes resulting from its relatively low bending modulus[17]. We demonstrate that the presence of out of plane ripples, whether they emerge from long wavelength fluctuations[18], or from edge stresses[19,20], leads to multiple orders of magnitude enhancement in the static non-reciprocity as compared to structures in which out of plane deformation do not occur. Our studies thus point to the potential of 2D materials as the basic building blocks of non-reciprocal nanoscale metamaterials.

**Non-Reciprocity of Planar Graphene Metamaterial**

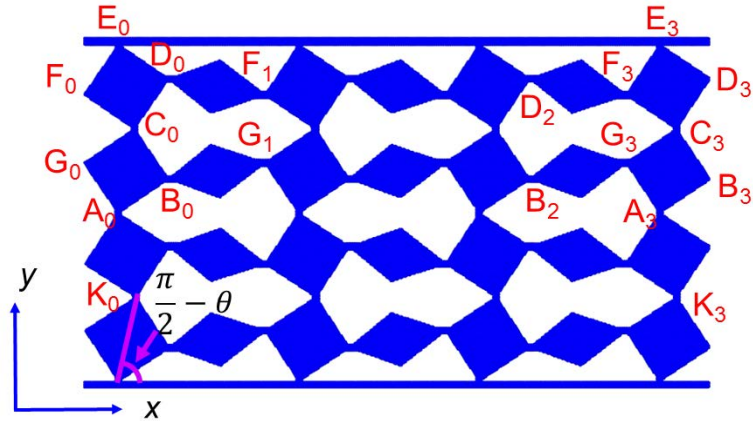

Figure 1: Schematic of graphene metamaterial, with different nodes labelled along with the asymmetry angle $\theta$.

The monolayer graphene metamaterial we consider in this work is illustrated in Figure 1. This graphene metamaterial is the 2D monolayer analog of the mechanical metamaterial structure considered by Coulais and co-workers in Figure 3 of their paper[10]. The metamaterial consists of diamond and square shaped monolayers of graphene that are connected via thin ligaments, and where the structural asymmetry is governed by the angle $\theta$. Molecular statics (MS) simulations with different techniques were used to investigate the non-reciprocity. In the first MS simulation approach, no perturbation along the out-of-plane direction is added, and we call this approach the 2D MS simulation. In the second MS simulation approach, the results of which are discussed in detail later, small random perturbations, which are used to induced out of plane rippling, are added to the out of plane (*z*)-displacements before any external loading is applied. We call this approach the perturbed MS simulation. These simulations will enable us to quantify the effect of intrinsic, out of plane ripples on the static non-reciprocity as compared to the planar 2D MS simulations. Details of the geometry of the graphene metamaterials and atomistic simulation methods can be seen in the Simulation

Methods section.

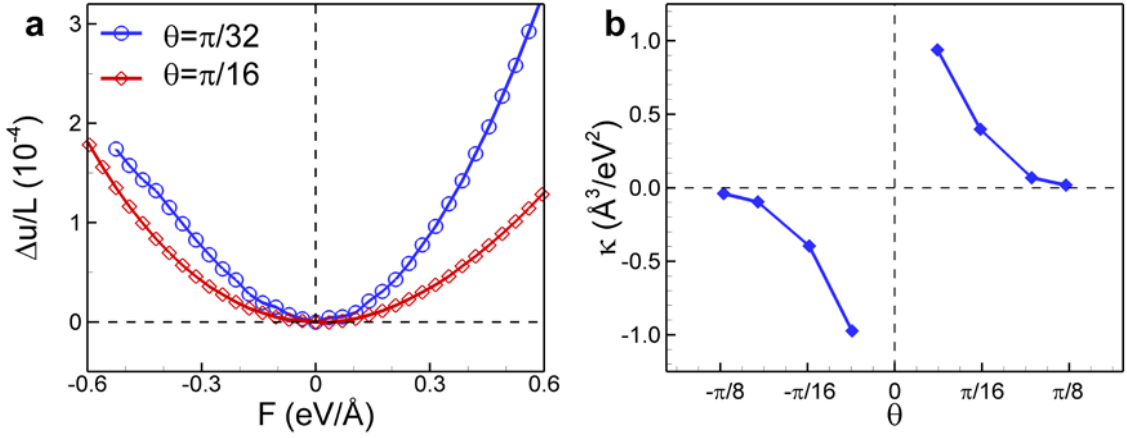

Figure 2: Non-reciprocity in the planar graphene metamaterials obtained by 2D MS simulation. (a): The change of the non-reciprocity parameter $\Delta u/L$ with applied force for different asymmetry angles where $L$ is the length of the graphene structure along the horizontal direction. (b): Non-reciprocity susceptibility parameter $\kappa = \Delta u/F^2$ versus the asymmetry angle.

We first discuss the 2D MS results, to highlight the static non-reciprocity of graphene when out of plane distortions do not occur, as in the bulk mechanical metamaterial studied by Coulais *et al*[10]. Similar to the work of Coulais *et al*., we define the non-reciprocity parameter to be $\Delta u = u_{0,R} + u_{3,L}$, where $u_{0,R}$ is the displacement of node $A_0$ due to the applied force at right end in Figure 1, while $u_{3,L}$ is the displacement at node $A_3$ due to the applied force at the left end. Figure S1 in the Supporting Information illustrates the structural response of the graphene metamaterial to an external force 1.4 eV/Å applied at the right end (node $A_3$ in Figure 1), and also at the left end (node $A_0$ in Figure 1). It is clear from Figure S1 that significantly larger deformation is observed when the force is applied at the right end than at the left end of the metamaterial, similar to the observation by Coulais *et al*.[10]. The displacement field for the graphene metamaterial in Figure S1 exhibits a large value at the right end, but with a large, nonlinear decay with distance away from the right end. Thus, for $\theta>0$, when we pull the structure from the right end, the displacement is large at the right end ($u_{3,R}$ is large) and it decreases significantly moving towards the left end. In contrast, the displacement at the left end node $A_0$ due to the force applied at the right end $u_{0,R}$ is much smaller than $u_{3,R}$. However, $u_{0,R}$ is still larger than $u_{3,L}$ and thus a non-reciprocal response of the graphene metamaterial is observed as demonstrated in Figure 2a.

Figure 2a quantifies the response of the graphene metamaterial for two different asymmetry angles, $\theta = \pi/32$ and $\theta = \pi/16$. For small forces, the relationship between the displacement and applied force is quadratic, $\Delta u = \kappa F^2$. Figure 2a demonstrates that the structure with $\theta = \pi/32$ is more asymmetric than the one with $\theta = \pi/16$, and also exhibits a larger displacement. This is shown more concretely by plotting the susceptibility parameter $\kappa$ as a function of asymmetry angle in Figure 2b, where we find, similar to the analysis of Coulais *et al.*, a divergence in the susceptibility parameter as the asymmetry angle $\theta \rightarrow 0$.

These 2D MS simulations serve to demonstrate that, like the bulk metamaterials studied by Coulais *et al.*[10], 2D graphene metamaterials with the same geometry and with the same in-plane deformations in response to the applied forces also exhibit static non-reciprocity, while also illustrating the potential of highly non-reciprocal graphene metamaterials for small asymmetry angles as illustrated in Figure 2b. However, the key question we wish to address in this work is this: what effects do physics that are unique or intrinsic to 2D materials, and that do not exist in bulk materials, have on the static non-reciprocity of graphene metamaterials? We examine this next using MS simulations with small perturbations in the out-of-plane direction.

**Non-Reciprocity of Rippled Graphene Metamaterial**

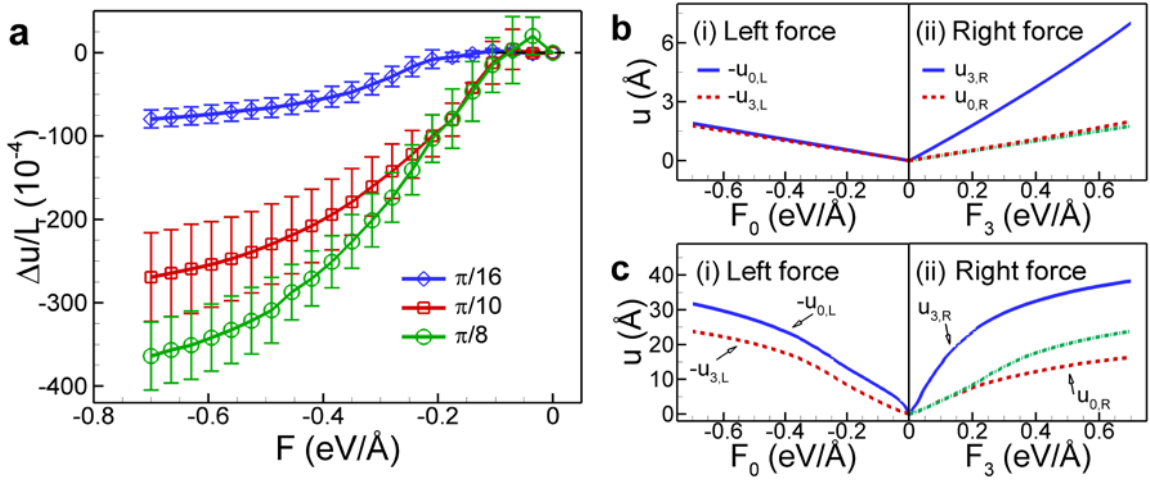

Figure 3: Non-reciprocity in the graphene metamaterials depicted in Figure 1 obtained by perturbed MS simulations. (a): The change of the non-reciprocity parameter of graphene metamaterial with applied force. The standard deviation was used to qualify the variation of the results. 2D MS simulations displacements at (b-i) nodes $A_0$ and $A_3$ due to forces applied at the left end (nodes $A_0$); (b-ii) nodes $A_0$ and $A_3$ due to forces applied at the right end (node $A_3$). Perturbed MS simulations displacements at (c-i) nodes $A_0$ and $A_3$ due to forces applied at the left end (node $A_0$); (c-ii) nodes $A_0$ and $A_3$ due to forces applied at the right end (node $A_3$). The green dashed and dotted lines in (b-ii) and (c-ii) are the mirror images of the red dash lines in (b-i) and (c-i), respectively; the line of refection is the vertical center line. For b-c, the asymmetry angle of $\theta = \pi/16$ was considered. Note the significantly enhanced displacements in (c) with rippling as compared to (b) without rippling.

We now examine the effects that out of plane ripples, which are intrinsic to graphene[18], have on the static non-reciprocity. To do so, we performed perturbed MS simulations, where small random displacements in the out of plane ($z$)-direction were applied to all atoms to mimic the effects of intrinsic rippling. Because of our interest in static, and not dynamic non-reciprocity, the small out of plane perturbations enabled us to generate out of plane ripples without the need for thermal fluctuations, which are a dynamic property. This further enabled us to isolate the effects of ripples on the static non-reciprocity via quasi-static MS simulations, without needing to utilize high strain-rate, dynamic molecular dynamics (MD) simulations.

The rippling is stochastic; 4 different perturbed MS simulations with different initial random perturbations in the $z$-direction were conducted, where we note that the energy of the equilibrium configuration of the graphene metamaterials with out of plane ripples is smaller than that of the flat graphene metamaterial without rippling, indicating that the rippled configuration is more energetically favorable than the flat configuration. Figure S2 in the Supporting Information demonstrates that a different rippling pattern will be observed throughout the metamaterial depending on the initial random perturbation. This is because different parts of the metamaterial can ripple in the positive or negative out of plane directions depending on the sign of the perturbation. However, we will demonstrate below that the stochastic nature of the rippling that is observed in the equilibrated structures in Figure S2 before forces are applied does not impact the results that we report.

Figure 3a shows the non-reciprocity parameter obtained by the perturbed MS simulations with out of plane rippling. In comparison to Figure 2a, where out of plane rippling was not considered, two important factors have changed. First, the non-reciprocity parameter is larger by two orders of magnitude when rippling is accounted for. Second, the sign of the non-reciprocity parameter when rippling is accounted for is negative, which is different from the flat graphene metamaterial in Figure 2a.

Figure 3 also presents the changes of the displacements at nodes $A_0$ and $A_3$ of the graphene metamaterial depicted in Figure 1 under loading obtained by the 2D MS simulations (Figure 3b) as well as perturbed MS simulations (Figure 3c). The perturbed MS displacements in Figure 3c are different from the 2D MS displacements in Figure 3b in two aspects. First, the displacements obtained by the perturbed MS simulations are much larger than those of the 2D MS simulations; we will mechanistically demonstrate later that this is due to the out of plane rippling. Second, the displacement is a nonlinear function of loading in the perturbed MS simulations in Figure 3c, which is different from the linear force-displacement relationship seen in the 2D MS simulations in Figure 3b. The linearity of the displacement curves in Figure 3b indicates that the stiffness of the 2D MS structures does not change with applied load. On the other hand, all curves of the perturbed MS results in Figure 3c are concave, i.e. the displacements increase with applied force but the incremental change decreases. This indicates that the stiffness of the rippled graphene metamaterials structures increases as the applied load increases.

Although the displacements in the perturbed MS simulations are larger than those of the 2D MS results, the degree of enhancement of the displacements (in the comparison with the 2D MS simulation) when the metamaterial is pulled at the right end is smaller than when the metamaterial is pulled from the left end. For example, as can be seen in Figure 3(b-c), for the applied force of magnitude 0.35 eV/Å, $|u_{3,L}|$ obtained by the perturbed MS simulation in Figure 3(c-i) is about 15 times larger than the corresponding 2D MS displacement in Figure 3(b-i) whereas $|u_{0,R}|$ obtained by the perturbed MS simulation is only about 10 times larger than the corresponding 2D MS displacement. Consequently, $|u_{0,R}|$ is smaller than $|u_{3,L}|$ leading to negative non-reciprocity parameter as shown in Figure 3a.

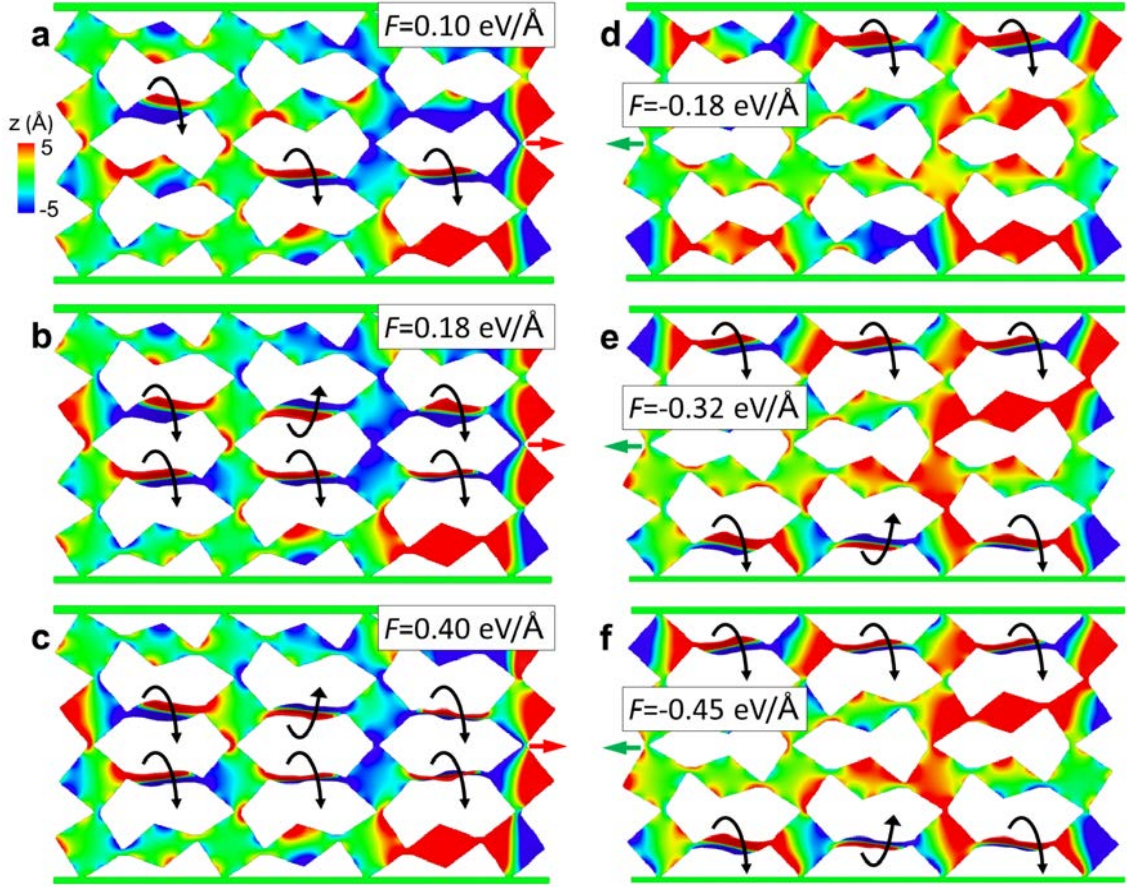

Figure 4: The configurations of the graphene metamaterial with the asymmetry angle $\theta = \frac{\pi}{16}$ under different applied forces at the right end (a-c) and at the left end (d-f).

Figure 4(a-c) presents the configurations of the rippled graphene metamaterial with the angle $\theta = \frac{\pi}{16}$ when it is pulled from the right end with different force magnitudes; the out of plane ($z$)-displacement magnitudes are shown. When the applied force increases in Figure 4(a-c), a notable displacement mode is observed (see supplemental movies). Specifically, for forces smaller than 0.18 eV/Å, the diamonds in the two center-rows rotate out of the $xy$-plane about the $x$-axis as shown in Figure 4a. After the force reaches 0.18 eV/Å, all diamond elements in the two middle rows have rotated out of the $xy$-plane, as shown in Figure 4b. As the force

increases beyond 0.18 eV/Å, the *y*-coordinates of the ligaments connecting the diamond elements to their square neighbors begin to align, with the alignment due to the applied tensile force essentially finishing when the force reaches 0.4 eV/Å as shown in Figure 4c.

We observed a strong correlation between the rotation of the diamonds and the displacements $u_{0,R}$ and $u_{3,R}$. Specifically, below a critical value of the applied force, the displacements $|u_{0,R}|$ and $|u_{3,R}|$ increase linearly as the applied force increases, as shown in Figure 3(c-ii). However, those displacements increase non-linearly when the applied forces exceeds about 0.18 eV/Å, which is when all diamonds in the center two rows have rotated out of the *xy*-plane. The stiffening of the rippled metamaterials in Figure 3(c-ii) occurs when the applied force exceeds 0.18 eV/Å because the displacement of the metamaterial shifts from the combination of in-plane stretching of the connecting ligaments and out of plane rotation of the center-row diamonds to primarily high energy in-plane stretching of all connecting ligaments.

The rotational displacement of the diamond elements is also observed in the rippled graphene metamaterial when pulled from the left end (see supplemental movies), though there are differences from the case of pulling from the right end that was just discussed. First, Figure 4(d-f) shows that it is the diamonds in the two boundary rows rather than the two center rows that rotate out of the *xy*-plane about the *x*-axis as the force increases. Second, the magnitude of the critical force at which all boundary-row diamonds have rotated out of the *xy*-plane is 0.32 eV/Å, which is larger than the critical force of 0.18 eV/Å when the rippled metamaterial is pulled from the right end. After the out of plane rotation of the diamond elements is completed, the metamaterial is again forced to deform via high energy stretching of the elements and ligaments, which increases the stiffness of the structure when the applied force exceeds the critical force. That explains why both displacements $|u_{0,L}|$ and $|u_{3,L}|$ in Figure 3(c-

i) increase linearly with force when the applied force is smaller than 0.32 eV/Å, and why both displacements increases non-linearly when the left end pulling force exceeds 0.32 eV/Å.

The rotational modes are what enable the significant increases in the magnitude of the non-reciprocal behavior for the graphene metamaterial with out of plane rippling. Figure 3a shows the non-reciprocity is two orders of magnitude larger than that for the graphene metamaterial that is constrained to remain planar in Figure 2a, where the 2D MS simulations of the planar graphene metamaterial showed similar static non-reciprocity to the macroscale experiments of Coulais *et al.*[10]. Specifically, Figure 3a demonstrates that the graphene metamaterials with out-of-plane rippling exhibit three distinct regimes of static non-reciprocity.

The first regime is for forces smaller than 0.18 eV/Å. Here, there is a smaller increase in non-reciprocity, though it is important to note that the value of the non-reciprocity parameter reaches nearly 100, which is still two orders of magnitude larger than the value of non-reciprocity of ~1 for the 2D planar graphene metamaterial shown in Figure 2a. This coincides, as shown in Figure 4(a-b), with the emergence of the diamond rotation mechanism. For this force range, both displacements $|u_{0,R}|$ and $|u_{3,L}|$ increase linearly with applied force as shown in Figure 3b, so the rate of increase in the non-reciprocity parameter is smaller than for the next regime.

The second regime is when the applied force is in the intermediate range (0.18eV/Å<$F$<0.32eV/Å). Here, the displacement $|u_{3,L}|$ continues to increase with applied force since the rotational mode in the case of pulling from the left continues to evolve as shown in Figures 4(d-e), whereas the displacement $|u_{0,R}|$ has a smaller increase because the rotational mechanism in the case of pulling from right has completed, as shown in Figure 4b. Therefore, the largest increases in non-reciprocity are observed in Figure 3a for this force range.

The final regime is for forces larger than about 0.32 eV/Å. As shown in Figure 4f, all metamaterial elements have rotated and aligned when pulling from the left, and thus in-plane stretching dominates and $|u_{3,L}|$ begins to increase more slowly, as shown in Figure 3a, leading to a decrease in the rate of increase in the non-reciprocity.

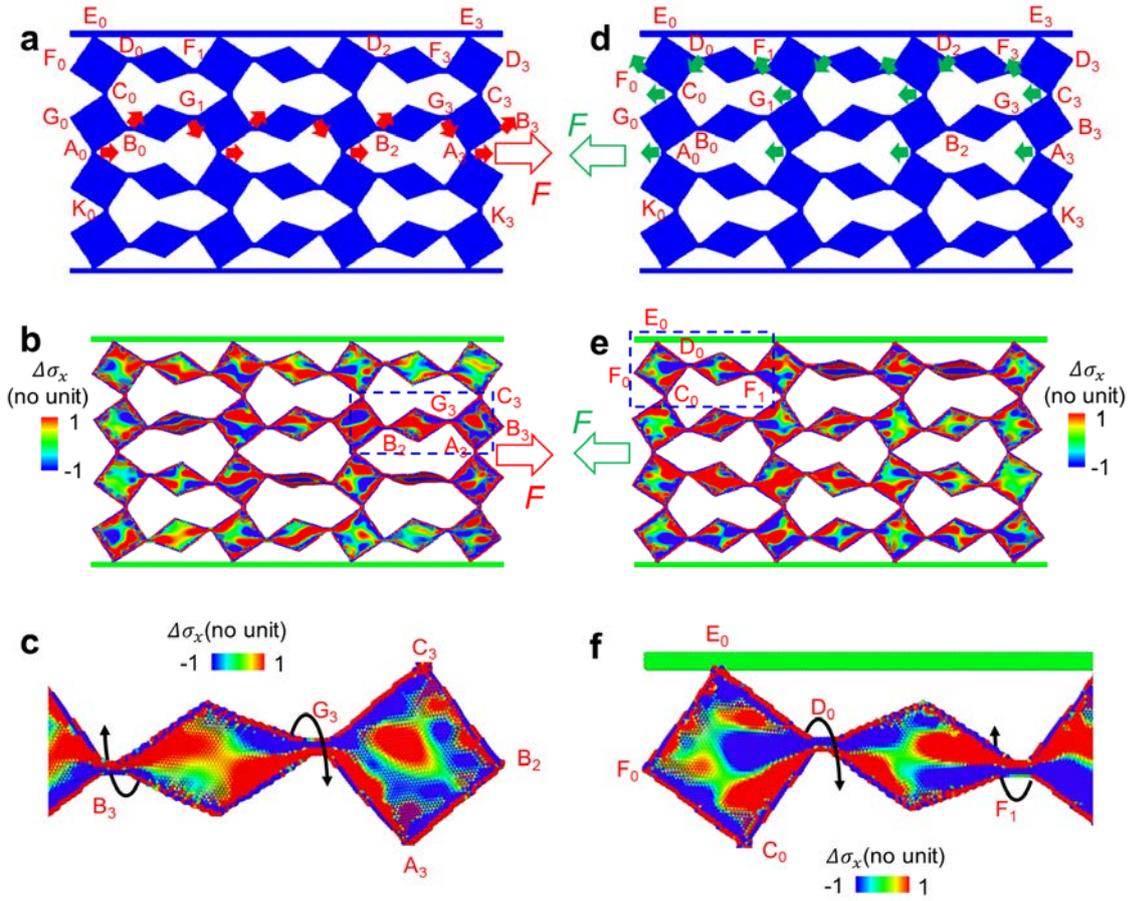

Figure 5: Displacement and stress fields resulting from applied force at right end (a-c) and from applied force at left end (d-f) when rippling is considered. (a,d): Displacements, and (b-c, e-f): Stress fields. Only displacements at some nodes are shown. For (b-c), the force $F$=0.11 eV/Å is applied, and for (e-f), the force $F$=-0.175 eV/Å is applied. Rotation of center row diamonds (for the case of applying force from right) and boundary diamonds (for the case of applying force from left) out of the $xy$-plane about the $x$-direction can be observed due

to the buckling of the ligaments which is due to the compressive axial stress in these ligaments. The asymmetry angle $\theta = \pi/16$ was considered.

The final question we address is to explain not only why the rotational mechanism occurs, but why the top and bottom rows exhibit the rotational mechanism when the metamaterial is pulled from the left in Figure 4 (d-f), whereas the middle two rows rotate when the metamaterial is pulled from the right in Figure 4 (a-c). The mechanism for pulling to the right is shown in Figures 5 (a-c), and S3. When the metamaterial is pulled to the right, a significant compressive stress is generated in the two center squares of the last column, as shown in Figure S3(a). This is because the two squares in rows 2 and 3 act as two bars connected at a hinge, where due to the angle of connection, the applied force to the right generates compression in both bars as shown in Figure S3(b). As shown in Figure S3(a), the compressive stress first generates an enhancement in the rippling amplitude, with part of the square element rippling in the positive $z$-direction, and the other part rippling down in the negative $z$-direction. The square element with nodes at $A_3B_3C_3G_3$ then rotates counterclockwise, which leads to a clockwise rotation of the diamond with nodes at $B_2$ and $G_3$ as shown in Figure 5a. Furthermore, as shown in Figures 5b and 5c, a portion of the ligament connecting square $A_3B_3C_3G_3$ and the diamond element is under compression, which causes out of plane buckling of that ligament portion. Because one portion of the ligament is under compression, leading to out of plane buckling, while the other part is in tension, rotation of the diamond element out of the $xy$-plane about the $x$-axis is observed with increasing force. The other connecting diamond elements in the center two rows exhibit a similar deformation transition as load is transferred to them from enhanced out of plane rippling to rotation of the element about the out of plane ($z$)-axis as force increases.

However, when force is applied to the left end, the rotational mechanism occurs in different rows of the metamaterial, as shown in Figure 5d. This is because a tensile stress is generated in the square elements as shown in Figure S4(a) due to the angle of connection shown in Figure S4(b). The tensile stress acts to flatten out the ripples as shown in Figure S4(a), and thus a rotation of the elements in the center two rows leading to rotation of the connecting diamond elements is not observed. However, the response of the boundary rows elements is different. Because the center rows move to the left when pulled at $A_0$, node $C_0$ in Figure 1 is also pulled to the left as shown in Figure 5d. This causes clockwise in-plane rotation of the top left square element $C_0F_0E_0D_0$. This rotation results in both compressive and tensile stresses in the ligament at node $D_0$ (Figures 5e and 5f). As the force increases, the compressive stress becomes sufficient to cause out of plane buckling of a portion of the connecting ligament, whereas the remainder of the ligament is under tension. This, as with the pulling to the right case, leads to rotation of the diamond element out of the *xy*-plane. As the force is increased, other diamond elements in the top and bottom rows exhibit the out of plane rotational mechanism; the square elements in those rows do not exhibit the rotational mechanism as they are more constrained by having three connected nodes, including one node fixed to the boundary rows.

**Conclusion**

We have used MS simulations to demonstrate that out of plane rippling, which is intrinsic to graphene, results in enhancements of the static non-reciprocity by two orders of magnitude as compared to graphene metamaterials that do not exhibit an out of plane deformation. In particular, the intrinsic rippling enables a unique, low energy rotational deformation in the graphene metamaterial. The asymmetry of the applied forces at the left and right ends that are needed to complete the rotational mechanism leads to a window of applied forces in which the static non-reciprocity increases significantly. The present results not only demonstrate the potential enhancements in non-reciprocity that can be enabled through out of plane deformations in atomically-thin 2D metamaterials, but also demonstrate the possibility of also achieving sign-tunable non-reciprocal behavior.

**Simulation Methods**

The monolayer graphene metamaterials consist of about 80000 carbon atoms, which were modeled using the second generation Rebo (REBO-II) potential, which has been shown to well-represent the large-strain mechanical behavior and properties of graphene[21]. We assigned the zigzag and armchair directions as the *x*- and *y*-directions, respectively. No periodic boundary conditions were applied in any direction. The top and bottom edges of the metamaterial were held fixed (see solid horizontal lines in Figure 1), while forces were applied to either node $A_0$ or node $A_3$ in Figure 1 in order to calculate the degree of static structural non-reciprocity. Molecular statics (MS) simulations were used to investigate the non-reciprocity, using the publicly-available simulation code LAMMPS[22].

As mentioned in the main text, we employed 2D MS simulations as well as perturbed MS simulations. The enhancements in non-reciprocity due to out of plane distortions were found by performing MS simulations with an initial out-of-plane perturbation. Specifically, the out of plane distortions, or rippling, are induced by adding small random perturbations to the out of plane (*z*)-displacements before any external loading is applied. The MS simulations correspond to a zero temperature (0 K) static simulation in which the equilibrium positions of the atoms in response to applied loading are found through energy minimization, and correspond to a situation in which thermal energy is absent from the system. The conjugate gradient method was used for all minimizations where a set of atomic positions in response to a specific external force is deemed to be convergent if the energy change between successive iterations divided by the energy magnitude is less than $10^{-14}$. In all MS simulations, before applying force, each structure was fully relaxed to attain the equilibrium state, after which forces were applied in increments of 0.005 eV/Å.

**Acknowledgements**

We gratefully acknowledge the support from the ICT R&D Program (No R0190-15-2012) of Institute for Information communications Technology Promotion (IITP) and from the Mid-Career Researcher Support Program (Grant No. 2014R1A2A2A09052374) of the National Research Foundation (NRF), which are funded by the MSIP of Korea. We also acknowledge with gratitude the PLSI supercomputing resources of the KISTI and the UNIST Supercomputing Center. HSP acknowledges the support of the Mechanical Engineering department at Boston University.